\documentclass[12pt]{article}
\usepackage{graphicx}

\usepackage{color}
\usepackage{setspace}
\usepackage{scicite}
 \usepackage{amsmath, amsthm, amssymb}

\usepackage{times}

\newenvironment{sciabstract}{%
\begin{quote} \bf}
{\end{quote}}

\newcounter{lastnote}
\newenvironment{scilastnote}{%
\setcounter{lastnote}{\value{enumiv}}%
\addtocounter{lastnote}{+1}%
\begin{list}%
{\arabic{lastnote}.}
{\setlength{\leftmargin}{.22in}}
{\setlength{\labelsep}{.5em}}}
{\end{list}}

\title{Sign-problem-free quantum Monte Carlo\\ of the onset of antiferromagnetism in metals$^\star$}

\author
{Erez Berg,$^{1,2\ast}$ Max A. Metlitski,$^{3}$ Subir Sachdev$^{1,4}$\\
\\
\normalsize{$^{1}$Department of Physics, Harvard University, Cambridge MA 02138, USA}\\
\normalsize{$^{2}$Department of Condensed Matter Physics, Weizmann Institute of Science, Rehovot 76100, Israel}\\
\normalsize{$^{3}$Kavli Institute for Theoretical Physics, University of California, Santa Barbara, CA 93106, USA}\\
\normalsize{$^{4}$Instituut-Lorentz for Theoretical Physics, Universiteit Leiden,}\\
\normalsize{P.O. Box 9506, 2300 RA Leiden, The Netherlands}\\
\\
\normalsize{$^\ast$To whom correspondence should be addressed; E-mail:  erez.berg@gmail.com.}\\
}

\date{}

\newcommand{\bk}{{\bf k}}
\newcommand{\bK}{{\bf K}}

\newcommand{\bq}{{\bf q}}

\newcommand{\beq}{\begin{equation}}
\newcommand{\eeq}{\end{equation}}
\def\bea{\begin{eqnarray}}
\def\eea{\end{eqnarray}}
\newcommand{\nn}{\nonumber \\}

\begin{document}


\baselineskip24pt


\maketitle


\begin{sciabstract}
The quantum theory of antiferromagnetism in metals is necessary
for our understanding of numerous intermetallic compounds of
widespread interest. In these systems, a quantum critical point
emerges as external parameters (such as chemical doping) are
varied. Because of the strong coupling nature of this critical
point, and the ``sign problem'' plaguing numerical quantum Monte
Carlo (QMC) methods, its theoretical understanding is still
incomplete. Here, we show that the universal low-energy theory for
the onset of antiferromagnetism in a metal can be realized in
lattice models, which are free from the sign problem and hence can
be simulated efficiently with QMC.
Our simulations show Fermi surface reconstruction and
unconventional spin-singlet superconductivity across the critical
point.
\end{sciabstract}

\begin{spacing}{0.7}
\noindent
{\footnotesize  $^\star$ This manuscript has been accepted for publication in Science. This version has not undergone final editing. Please refer to the complete version of record at http://www.sciencemag.org/. The manuscript may not be reproduced or used in any manner that does not fall within the fair use provisions of the Copyright Act without the prior, written permission of AAAS.
}
\end{spacing}


The presence of an antiferromagnetic transition in a metal is
common to compounds such as electron-doped
cuprates\cite{Kartsov2}, iron based superconductors\cite{matsuda},
and heavy fermion Kondo lattice systems\cite{park}. Whereas our
understanding of quantum antiferromagnetism in insulators has seen
remarkable advances \cite{balents}, analogous problems in metals
are far more complicated because of the subtle interplay between
the low energy fermionic quasiparticles on the Fermi surface, and
the quantum fluctuations of the antiferromagnetic order parameter.
In addition, the presence of the Fermi surface has hampered large
scale numerical studies, because QMC algorithms are afflicted by
the well-known fermion sign problem. Such algorithms express the
partition function as a sum over Feynman histories, and the sign
problem arises when the weights assigned to the trajectories are
not all positive because of quantum interference effects. A
general solution to the fermion sign problem has been proved to be
in the computational complexity class of nondeterministic
polynomial (NP) hard \cite{troyer}, and so there has been little
hope that the antiferromagnetic quantum critical point could be
elucidated by computational studies.

Application of the methods of quantum field theory and the
renormalization group to the onset of antiferromagnetism in a
metal\cite{hertz}, has identified \cite{AC,AC1} a universal
quantum field theory which captures all the singular low energy
quantum fluctuations that control the quantum critical point and
deviations from the Fermi liquid physics of traditional metals.
In two spatial dimensions, the field theory is expressed in terms of fermionic excitations in
the vicinity of a finite number of `hot spots' on the Fermi
surface, and is thus independent of the details of the fermionic
band-structure, except for the number of hot-spots and
Fermi-velocities at the hot-spots\cite{SM}. Recent work
\cite{maxsdw,maxsdw1} has shown that the renormalization group and
Feynman graph expansions of the field theory flow to strong
coupling, making further analytical
progress difficult.

Here, we show that the universal quantum field theory can be
realized in lattice models which are free of the sign problem, and
so is amenable to large scale QMC studies.
Our claim does not contradict the no-go theorem of
Ref.~\cite{troyer}, because we do not provide a general recipe for
eliminating the sign problem. However, we will eliminate it for
the specific case of the onset of antiferromagnetic order in a
two-dimensional metal, provided the perturbative arguments on the
importance of the hot spots to the quantum field theory
\cite{AC,AC1,maxsdw,maxsdw1} apply. Our modified lattice model has
at least two bands. Therefore, in cases in which there is only a
single active band at the transition, such as in the
electron-doped cuprates, our method requires modifying Fermi
surface far away from the hot spots; we show that this can be done
while preserving the universal low-energy properties of the
antiferromagnetic critical point. On the other hand, our method
applies to multi-band situations (such as in the iron-based
superconductors) without changes to their Fermi surface
configuration. Being a low-energy effective theory, the method
will not apply where the proximity of a Mott insulator is
important, as is likely the case in the hole-doped
cuprates\cite{doiron,doiron1,kotliar,tremblay,millis}.

\begin{figure}
  \centering
  \includegraphics[width=7in]{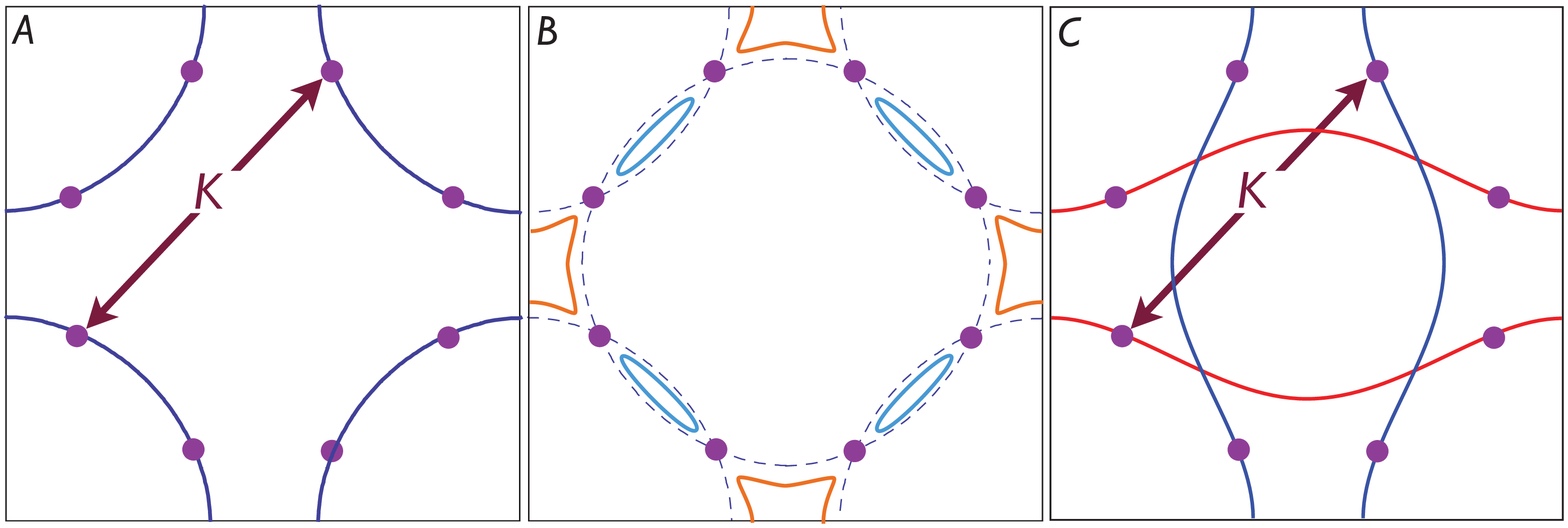}
  \caption{(\emph{A}) Fermi surface of the Fermi liquid phase of a single band model on the square lattice with unit lattice spacing.
  The ``hot spots'' are denoted by the filled circles. 
  (\emph{B}) The reconstructed Fermi surface in the metal with SDW order. The dashed lines show the Fermi surface in the metal without SDW order, and its
  translation by $\bK$.
  Gaps have opened at the hot spots, leading to small ``pocket'' Fermi surfaces.
  ({\em C\/}) A deformed Fermi surface of the metal without SDW order, in which the vicinities of the hot spots are unchanged from (A). The horizontal and vertical Fermi surfaces now belong to separate electronic bands.}
  \label{fig:deform}
\end{figure}

To illustrate our method, we now consider the onset of
antiferromagnetic order in a simple one-band model on the square
lattice, as is appropriate for the electron-doped cuprates. The
electrons, $c_{\bk}$ (the spin index is left implicit), with
dispersion $\varepsilon_\bk$, have a single ``large'' Fermi
surface (Fig~\ref{fig:deform}A). The antiferromagnetic order
parameter is $\vec{\varphi}_\bq$; we will assume the important
fluctuations of $\vec{\varphi}_\bq$ are restricted to small values
of $|\bq|$, much smaller than the size of the Brillouin zone. The
antiferromagnetic ordering wavevector is $\bK = (\pi,\pi)$, and
$\vec{\varphi}_\bq$ represents the electron spin density at the
wavevector $\bK + \bq$; we will also refer to the
antiferromagnetic order as spin density wave (SDW) order. We can
now write the electron part of the Hamiltonian as \beq H =
\sum_\bk \varepsilon_{\bk} \, c_\bk^\dagger
c^{\vphantom{\dagger}}_\bk + \lambda \sum_{\bk,\bq} c^\dagger_{\bk
+ \bK + \bq} \, ( \vec{s} \cdot \vec{\varphi}_\bq ) \,
c^{\vphantom{\dagger}}_\bk \label{hsdw} \eeq where $\lambda$ is
the `Yukawa' coupling between the electrons and the SDW order, and
$\vec{s}$ are the Pauli matrices. The Yukawa term is the simplest
coupling consistent with translational symmetry and spin-rotation
invariance, and can be derived {\em e.g.} by decoupling of the
repulsive interaction in a Hubbard model by an auxiliary field
which maps to $\vec{\varphi}$ in the long-wavelength
limit\cite{AbanovAdvPhys}. The hot spots are at $\bk$ for which
$\varepsilon_\bk = \varepsilon_{\bk + \bK} = 0$
(Fig.~\ref{fig:deform}A); at these points, $\vec{\varphi}_{\bq =
0}$ scatters electrons between initial and final states which are
both on the Fermi surface. To obtain the electron Fermi surface in
a metal with SDW order, we replace $\vec{\varphi}_\bq$ by its
expectation value $\langle \vec{\varphi}_q \rangle = \vec{N}
\delta_{\bq,0}$ (where $\vec{N}$ is the staggered magnetization),
and recompute the electron dispersion; this leads to the Fermi
surface reconstruction shown in Fig.~\ref{fig:deform}B.

We now describe our method to deform the model, such that the sign
problem is avoided, while preserving the structure of the hot
spots. 
Let us separate the hot spots into two groups, so that $\bK$ only
connects hot spots from one group to the other. Now deform the
one-band electronic dispersion to a two-band model with an
additional `orbital' label, so that all the hot spots in one group
are on the Fermi surfaces of the first band, while the hot spots
of the other group reside on the Fermi surfaces of the second band
(an example of such a dispersion is shown in
Fig.~\ref{fig:deform}C, in which the `horizontal' and `vertical'
Fermi surfaces are part of two separate electronic bands). Note
that the vicinities of the hot spots in the two-band model are
essentially identical to those in the one-band model in
Fig.~\ref{fig:deform}A, and so the same low energy theory for the onset
of antiferromagnetism applies to both models. With no further assumptions, the deformed
model has only positive weights in a suitable quantum Monte Carlo
realization.

\begin{figure}
\centerline{\includegraphics[width=5in]{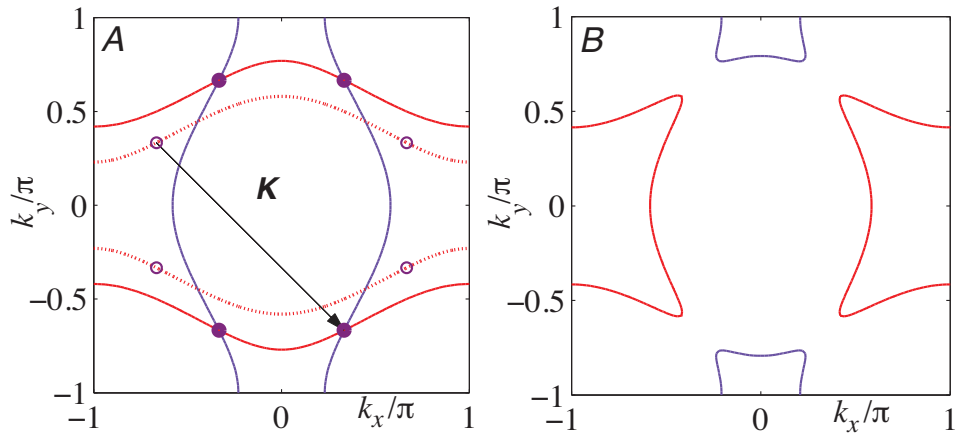}}
\caption{({\em A\/}) Fermi surfaces (full lines) of $L_F$ for free
$\psi_{x,y}$ fermions with the parameters listed in the text. The
dashed lines show the portion of the Fermi surface in
Fig.~\ref{fig:deform}c which was shifted by $\bK$ to obtain the
$\psi_y$ Fermi surface. Now the hot spots are at the intersections
of the Fermi surfaces. ({\em B\/}) Mean-field $\psi_{x,y}$ Fermi
surfaces with SDW order $|\langle \vec{\varphi} \rangle| = 0.25$.}
\label{fig:FS}
\end{figure}

We will write down a specific lattice model for which we will
establish a sign-free Monte Carlo algorithm, and then present
numerical results. We begin with the band structure of the $c_\bk$
electrons in Fig.~\ref{fig:deform}C. We write the band with
vertical Fermi surfaces in terms of fermions $\psi_x$ with $c_\bk
\rightarrow \psi_{x, \bk}$, and the band with horizontal Fermi
surfaces in terms of fermions $\psi_y$ with $c_\bk \rightarrow
\psi_{y, \bk + \bK}$. This leads to the $\psi_{x,y}$ Fermi
surfaces shown in Fig.~\ref{fig:FS}A. Then our model has the
action $S = S_F + S_\varphi = \int_0^\beta d \tau (L_F +
L_\varphi)$ with \bea L_{F}\!\! &=& \!\!\!\!\!\
\sum_{i,j,\alpha=x,y}\psi_{\alpha
i}^{\dagger}\left[\left(\partial_{\tau}-\mu\right)\delta_{ij}-t_{\alpha,ij}\right]\psi_{\alpha
j}
 +\lambda\sum_{i}
\psi_{xi}^{\dagger}\left(\vec{s}\cdot\vec{\varphi}_{i}\right)\psi_{yi}+H.c.,
\nn L_{\varphi} &=&
\frac{1}{2}\sum_{i}\frac{1}{c^{2}}\left(\frac{d\vec{\varphi}_{i}}{d\tau}\right)^{2}+\frac{1}{2}\sum_{\left\langle
i,j\right\rangle
}\left(\vec{\varphi}_{i}-\vec{\varphi}_{j}\right)^{2}+\sum_{i}\left(\frac{r}{2}\vec{\varphi}_{i}^{2}+\frac{u}{4}(\vec{\varphi}_{i}^2)^{2}\right).
\label{Z2} \eea Here $i,j$ run over the sites of the square
lattice, $\tau$ is the imaginary time and $\beta$ - the inverse
temperature. The parameter $r$ will be used to tune across the
quantum critical point, and $u$ is a non-linear self-coupling of
$\vec{\varphi}$. The $\psi_x$ ($\psi_y$) fermion hops along the
horizontal (vertical) direction with an amplitude
$t_{\parallel}=-1$ $(+1)$, and along the vertical (horizontal)
direction with an amplitude $t_{\perp}= -0.5$ $(0.5)$,
respectively; the resulting band structure is shown in
Fig.~\ref{fig:FS}A (solid lines). The model has $C_4$ symmetry,
and its apparent violation is an artifact of the shifting of the $\psi_y$
fermions by $\bK$. We chose the chemical potential $\mu_{1}
=\mu_{2}=-0.5$, $c =1$, $u=1$, and $\lambda=1$.

By construction, the modified two-band model has the same hot spot
structure as the original one-band model. Therefore, we argue that
it preserves the universal properties of the antiferromagnetic
transition. We prove\cite{SM} that the introduction of the second
band eliminates the sign problem in this model.

Note that it is possibly to analytically integrate out $\vec{\varphi}$ in Eq.~(\ref{Z2}),
and establish equivalence to a large class of Hubbard-like models to which our method applies. However, we choose
to keep $\vec{\varphi}$ as in independent degree of freedom because it keeps the physics transparent
and streamlines the analysis.

We have performed determinant Monte Carlo simulations of the
action (\ref{Z2}) using the algorithm described in Refs.
\cite{Blankenbecler1981,White1989,Assaad2008}, for systems of
linear size up to $L=14$ and inverse temperature $\beta=14$, with
either periodic or anti-periodic boundary conditions. An imaginary
time step of $\Delta \tau=0.1$ was used in most of the
calculations; we checked that the results do not change for
$\Delta \tau=0.05$. Up to $50000$ Monte Carlo sweeps were
performed for each run, giving a statistical error for most
measured quantities of a few percent.

\begin{figure}
\centerline{\includegraphics[width=7.5in]{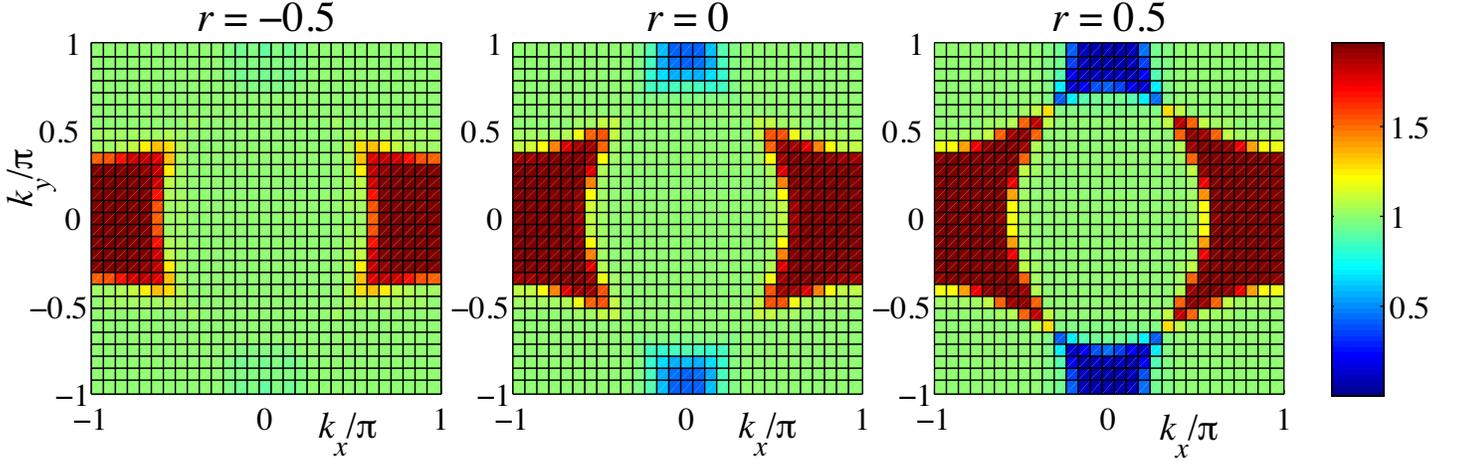}}
\caption{Quantum Monte-Carlo results for the fermion occupation
number $n_\bk = \langle \psi^\dagger_{x\bk} \psi_{x\bk} +
\psi^\dagger_{y\bk} \psi_{y\bk}\rangle/2 $ as a function of $\bk$
across the Brillouin zone, for systems with $L=14$, $\beta=14$,
and $r=-0.5,0,0.5$. In order to enhance the resolution, results
from simulations with either periodic or anti-periodic boundary
conditions in the $x$ and $y$ directions were combined. Despite
appearances, full square lattice symmetry is preserved in all our
computations for the original $c_\bk$ fermions.} \label{fig:nk}
\end{figure}

First, we present results showing the reconstruction of the Fermi
surface across the SDW transition. Fig. \ref{fig:nk} shows the
fermion occupation number summed over the two flavors of fermions
as a function of quasi-momentum. The Fermi surfaces are clearly
visible as discontinuities. $r=0.5$ is found to be on the
disordered side of the SDW critical point, and the Fermi surface
closely resembles the one in Fig. \ref{fig:FS}A. At $r=0$, a gap
opens at the hot spots, and the Fermi surface is reconstructed
into electron and hole pockets, as in the SDW ordered state in
Fig.~\ref{fig:FS}b. Decreasing $r$ further to $-0.5$ increases the
magnitude of the SDW order parameter, and causes the hole pockets
to disappear and the electron pockets to shrink.

\begin{figure}
\centerline{\includegraphics[width=7in]{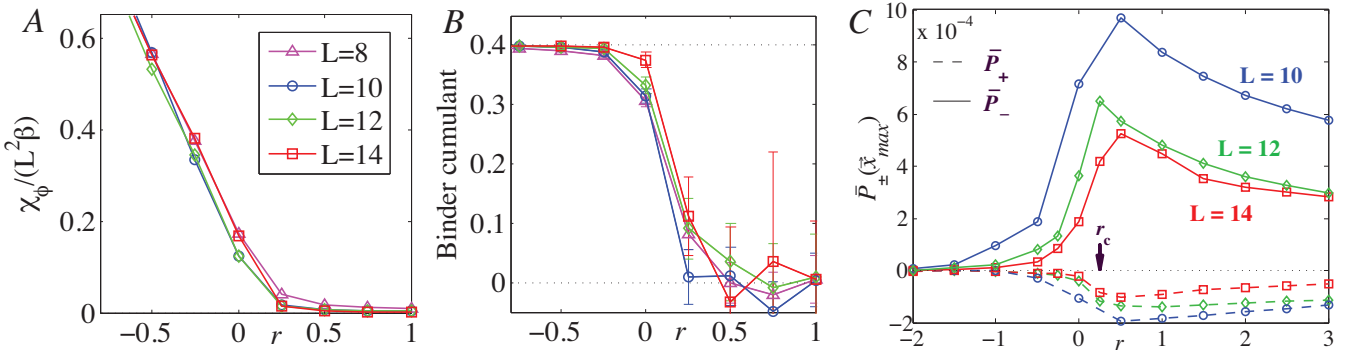}}
\caption{(\emph{A}) The SDW susceptibility $\chi_\varphi$,
normalized by $L^2\beta$, as a function of $r$, for systems of
size $L=8,10,12,14$ and $\beta=L$ for each curve. The
statistical errors in $\chi_\phi$ are smaller than the symbol
size. (\emph{B}) The Binder cumulant for an O(3) order parameter
$C_B = 1 - \frac{3\langle {\vec \Phi}^4 \rangle}{5\langle{\vec
\Phi}^2\rangle^2}$, where ${\vec \Phi} =
\frac{1}{\mathcal{N}}\sum_i {\vec \varphi}_i$, approaching the
expected values of $0.4$ and $0$ in the two phases. (\emph{C})
Equal-time pairing correlations in systems of size $L=10,12,14$
and $\beta=L$ for each curve, as a function of $r$. Dashed (solid) lines
show $\bar{P}_+$ ($\bar{P}_-$), corresponding to $A_{1g}$
($B_{1g}$) superconducting order parameters, in which the pairing
amplitude in the two fermion flavors is of the same (opposite)
sign, respectively. $r_c$ is the estimated position of the SDW
critical point.} \label{fig:chi}
\end{figure}

To examine the magnetic transition, we computed the SDW
susceptibility $ \chi_\varphi = \sum_i \int_0^\beta d\tau \langle
\vec{\varphi}_i(\tau)\cdot \vec{\varphi}_0(0) \rangle.$
Figure~\ref{fig:chi}A shows $\chi_\varphi$ normalized by $L^2
\beta$ as a function of $r$. In order to extract information about
the zero-temperature limit, we scale $\beta$ with the linear
system size; in the appropriate units, $\beta=L$ was used. We
observe a rapid upturn in $\chi_\varphi$ near $r=0.25$. For
$r<0.25$, $\chi_\varphi/(L^2 \beta)$ for different system sizes
and inverse temperatures nearly collapse on top of each other,
which is the expected behavior on the ordered side of the
transition. The results are consistent with a second-order
transition 
at $r_c\approx 0.25$. 
This is further supported by the Binder cumulant in Fig.
\ref{fig:chi}B, where we observe the expected behavior in both
phases, separated by a critical point at $r_c = 0.25 \pm 0.1$.

The SDW critical modes mediate effective inter-fermion
interactions, which can lead to instabilities of the Fermi
surface. As a result, additional competing phases can appear. Near
the SDW critical point, these instabilities are a result of a
subtle competition between the enhancement of the SDW
fluctuations, which tends to strengthen the effective
interactions, and the loss of coherence of the fermionic
quasi-particles\cite{maxsdw,maxsdw1}. Superconductivity is a
natural candidate for the leading potential instability. In order
to examine the emergence of a superconducting phase near the SDW
critical point, we have computed equal-time pairing correlations
$P_{\pm}(\vec{x}_i) = \langle
\Delta^{\vphantom{\dagger}}_{\pm}({\vec{x}}_i)
\Delta_{\pm}(0)^\dagger \rangle.$ Here, $\Delta_{\pm}({\vec{x}}_i)
= is^y_{ab} (\psi_{ix a}\psi_{ix b} \pm \psi_{iy a}\psi_{iy b})$
(where $a,b=\uparrow,\downarrow$ are spin indices) are
superconducting order parameters with either a $+$ or $-$ relative
sign between the two fermionic flavors (square lattice symmetry
$A_{1g}$ and $B_{1g}$, respectively).

In order to probe for long-range order, we measured
$P_{\pm}(\vec{x}_i)$ near the maximum range
$\vec{x}_{max}=(L/2,L/2)$. We plot $\bar{P}_\pm
(\vec{x}_{max})=\frac{1}{9}\sum^1_{\epsilon_{x,y}=-1} P_\pm
(\vec{x}_{max}+\epsilon_x {\vec \eta}_x + \epsilon_y {\vec
\eta}_y)$, where ${\vec \eta}_x = (1,0)$ and ${\vec \eta}_y =
(0,1)$, in Fig. \ref{fig:chi}C. Long-range superconducting order
at $\beta \rightarrow \infty$ would correspond to superconducting
correlations that saturate to a constant upon increasing $L$ and
$\beta$.

The $B_{1g}$ pairing correlations are found to be significantly
enhanced in the vicinity of the SDW critical point, $r_c\approx
0.25$. The $A_{1g}$ correlations are significantly smaller in
magnitude and negative in sign. This is consistent with the
expectation that the effective attraction mediated by magnetic
fluctuations promotes superconductivity with a sign change between
the two orbitals\cite{Scalapino1986,Monthoux1991}. 

The maximum of the $B_{1g}$ correlations occurs for $r\approx
0.5$, on the disordered side of the magnetic critical point which
is located at $r_c\approx 0.25$~\cite{moon}. Interestingly, the
suppression of the superconducting correlations away from the
optimal $r$ is very asymmetric: whereas the pairing correlations
decrease gradually for $r>r_c$, they are suppressed dramatically
for $r < r_c$. This may be a result of the opening of an SDW gap
on portions of the Fermi surface. 

The method described in this Letter opens the way to study various
physical aspects of spin density wave transitions in metals, in a
numerically exact way. The interplay between unconventional
superconductivity and magnetism and possible non-Fermi liquid
behavior in the quantum critical regime should now be accessible.
Moreover, such simulations will provide controlled benchmarks for
analytic approximations\cite{AC,AC1,maxsdw,maxsdw1}.

The two-band model presented here is a member of a wider family of
strongly correlated fermionic models that can be rendered free of
the sign problem. It has already been known that some models with
two flavors of fermions interacting via a four-fermion interaction
are sign problem free at generic fermion
density~\cite{Motome1997}. Remarkably, these models do not rely on
any specific characteristic of the electron dispersion; e.g. there
is no requirement for particle-hole symmetry, or any symmetry that
relates the two bands. Extensions of this trick to related models
of physical interest should be possible. 

\clearpage

\clearpage

\begin{center}
{\LARGE  Supplementary Material}
\end{center}
\section*{Quantum Field Theory}

In this section, we briefly review the universal field theory for
the SDW transition developed in Refs.~(\textit{7,8,10,11}). 
This
effective field theory has been analyzed perturbatively in the
order parameter-fermion coupling $\lambda$.

One of the key results of the analysis is that, if we ignore the
possibility of a high-energy pairing instability, the fermion and
order parameter propagators acquire \emph{universal} singular
parts which depend only on the structure of the hot spots (e.g.
the velocities at the hot spots and the angle between them). This
justifies the assumptions behind the construction of the lattice
model presented in the main text: as long as the structure of the
hot spots is preserved, we expect the universal behavior near the
antiferromagnetic critical point to be unchanged. The microscopic
parameters of the model should only come in through the
ultraviolet cutoffs to the critical fluctuations. These cutoffs
can, in principle, be set by matching at high energy scales.
Specifically, we can match a sign-problem-free lattice model to a
Hubbard-like model by equating their hot-spot Fermi surfaces and
Fermi velocities. Other parameters of the sign-problem-free model
can be determined by matching its physical observables to those of
the sign-problem-present Hubbard model at temperatures high enough
to allow accurate computations by other methods for the latter
model. The sign-problem-free model can then be used to compute
observables at low temperatures.

Another observation made in Ref.~(\textit{10}), however, is that
ultimately the conventional ways to control the perturbative
series for the effective field theory, such as an expansion in the
inverse number of fermion flavors, are uncontrolled for this
problem. The fate of the flow to strong coupling has to be
resolved by numerical simulations. For this we need a lattice
regularization of the continuum quantum field theory, and the
lattice model considered in the main body of the paper provides
precisely such a regularization.

The field theory is formulated in terms of the fermion excitations
in the vicinity of the hot spots. The antiferromagnetic order
parameter $\vec{\varphi}$, with wavevector $\bK$, connects
fermions at a hot spot at wavevector $\bk$ with fermions at a hot
spot with wavevector $\bk + \bK$; both fermions are on the Fermi
surface if $\varepsilon_{\bk} = \varepsilon_{\bk + \bK} = 0$, and
this defines the allowed values of $\bk$. Linearizing the fermion
dispersion about the hot spots, and expanding the order parameter
in spatial gradients, in two spatial dimensions we obtain the
Lagrangian $L = L_\psi + L_\varphi$ where \begin{align*}
\mathcal{L}_\psi &= \psi_{1}^{ \dagger} \left(
\partial_\tau - i {\bf v}_{1} \cdot {\bf \nabla} \right) \psi_{1}
+ \psi_{2}^{ \dagger}\left(  \partial_\tau - i {\bf v}_{2} \cdot
{\bf \nabla} \right) \psi_{2 }  + \lambda \vec{\varphi} \cdot (
\psi_1^{ \dagger} \vec{s} \psi_2 + \mbox{H.c.} ) \nn
\mathcal{L}_\varphi &= \frac{1}{2 c^2} (\partial_\tau
\vec{\varphi} )^2 + \frac{1}{2} ({\bf \nabla} \vec{\varphi})^2 +
\frac{r}{2} \vec{\varphi}^2 + \frac{u}{4} ( \vec{\varphi}^2 )^2
 \label{qft} \tag{S1}    \end{align*}
Here $\psi_a$, with $a=1,2$, are two species of low energy
fermions in the vicinity of the hot spots at $\bk$ and $\bk +
\bK$, and ${\bf v}_{a}$ are their Fermi velocities. A similar
Lagrangian applies to the other hot spots. This theory has the
same general structure as coupled fermion-boson theory in particle
physics, such as the Gross-Neveu model~(\textit{25}), with 
fermions and bosons coupled via trilinear ``Yukawa'' coupling
$\lambda$. The key difference is in the fermion dispersion, which
does not have a relativistic form. In the relativistic cases, the
fermion dispersion has a Dirac form with energy $\sim \pm v |{\bf
k}|$, and this vanishes only at isolated points in the Brilluoin
zone. The resulting fermion-boson theory is well understood
~(\textit{25}). In our case, the fermions dispersion $\sim {\bf v}
\cdot {\bf k}$, and this vanishes on a {\em line} in the Brillouin
zone which is orthogonal to ${\bf v}$. This is the central
difference which makes the quantum field theory in Eq.~(\ref{qft})
strongly coupled.

Let us parametrize the Fermi velocities by \beq {\bf v}_1 = (v_x,
v_y) \,, \qquad {\bf v}_2 = (- v_x, v_y) \,. \tag{S2} \eeq Here,
for convenience, we have rotated the coordinates by $45^\circ$
relative to Fig. 1a in the main text. It is useful to introduce
the ratio and the modulus \beq
 \tan \phi \equiv \frac{v_y}{v_x} \,, \qquad v = |{\bf v}| \,.
 \tag{S3}
\eeq Here $0 < 2 \phi < \pi$ is the angle between the Fermi
surfaces at the hot spot. We now present the singular terms in the
low energy spectrum at two loop order, as obtained in
Refs.~(\textit{10,17}). We will restrict the expressions to
precisely at the quantum critical point at zero temperature; all
our conclusions, and similar but lengthier expressions, apply also
close to the quantum critical point and at low temperatures. For
the fermion Green's function we have \beq G^{-1}_a(\omega, {\bf
p}) = -{\bf v}_a \cdot {\bf p} + \frac{3 v \sin 2 \phi}{8} i \,
{\rm sgn}(\omega) \left(\sqrt{\gamma |\omega| + \frac{({\bf
v}_{\bar{a}} \cdot {\bf p})^2}{v^2}} - \frac{|{\bf v}_{\bar{a}}
\cdot {\bf p}|}{v} \right) \,, \label{eq:fermionG0}\tag{S4}\eeq
where $\omega$ is an imaginary frequency, $\bar{1} = 2$ and
$\bar{2} = 1$, and \beq\label{eq:gam} \gamma = \frac{N_h
\lambda^2}{2\pi v_x v_y} \,, \tag{S5}\eeq where $N_h$ is the
number of pairs of hot spots ($N_h = 4$ for the electron-doped
cuprates). In the expression (\ref{eq:fermionG0}) we have dropped
the bare free fermion $i \omega$ term because it is not as
singular as the self-energy correction from the fluctuations of
the antiferromagnetic order, and we have not explicitly written
the real part of the self energy which renormalizes the velocities
$v_x$ and $v_y$. The singular part of the propagator of the boson
$\vec{\varphi}$ is \beq\label{eq:bosonD} D^{-1}(\omega ,\vec p \,)
= \gamma |\omega| + {\bf p}^{\,2}\,. \tag{S6}\eeq

In these expressions above the strength of the interactions is
controlled by the Yukawa coupling $\lambda$, and hence via the
value of $\gamma$. However a key observation is that dependence on
$\lambda$ can be scaled away, and the above low energy spectra are
actually universal. Indeed, it is easily seen from
Eqs.~(\ref{eq:fermionG0},\ref{eq:bosonD}) that after rescaling
momenta by ${\bf p} \rightarrow \lambda {\bf p}$, the $\lambda$
dependence appears as overall prefactors which can be absorbed
into a rescaling of the fields. This independence on the value of
$\lambda$ is a general feature of the low-energy quantum field
theory (\textit{10}), and is crucial to its properties. One of its
consequences appeared in the leading log estimate of the pairing
instability presented in Ref.~(\textit{10}), which was found to be
a logarithm-squared term with a $\lambda$-independent prefactor.


\section*{Quantum Monte Carlo}

We set up $S_F+S_\varphi$ for a Monte Carlo
study (\textit{17}). 
Discretizing imaginary time, the
partition function becomes
\begin{equation*}
Z=\int d\varphi\exp\left(-S_{\varphi}\right){\rm
Tr}_\psi\left[\prod_{n=1}^{N}\hat{B}_{n}\right]+O\left(\Delta\tau^{2}\right),\label{Z1}\tag{S7}
\end{equation*}
where $\Delta\tau$ is an imaginary time step, $\beta=N\Delta\tau$,
and the operators $\hat{B}_{n}$ are given by
\begin{equation*}
\hat{B}_{n}=e^{-\frac{1}{2}\Delta\tau\psi^{\dagger}K\psi}e^{-\Delta\tau\psi^{\dagger}V_{n}\psi}e^{-\frac{1}{2}\Delta\tau\psi^{\dagger}K\psi}.\tag{S8}
\end{equation*}
$K$ and $V_{n}$ are matrices given by
\[
K_{i,j;\alpha,\alpha^{\prime};s,s^{\prime}}=\delta_{s,s^{\prime}}\delta_{\alpha,\alpha^{\prime}}\left(-t_{\alpha,ij}-\mu\right)
\]
\begin{equation*}
V_{n;i,j;\alpha,\alpha^{\prime};s,s^{\prime}}=\lambda\left(\sigma_{1}\right)_{\alpha,\alpha^{\prime}}\delta_{ij}
\left[\vec{s}\cdot\vec{\varphi}_{i}\left(n\Delta\tau\right)\right]_{s,s^{\prime}},\tag{S9}
\end{equation*}
where $i,j$ are spatial indices, $\sigma_{1}$ is a Pauli matrix,
$\alpha,\alpha^{\prime}=x,y$ are flavor indices, and
$s,s^{\prime}=\uparrow,\downarrow$ are spin indices.
$\psi^{\dagger}$ is a vector of fermionic operators,
\begin{align*}
\psi^{\dagger}=\left(\psi_{x,1,\uparrow}^{\dagger},\dots,\psi_{x,\mathcal{N},\uparrow}^{\dagger},\psi_{x,1,\downarrow}^{\dagger},\dots,\psi_{x,\mathcal{N},\downarrow}^{\dagger},
\right. \nn  \quad \quad \quad \left.
\psi_{y,1,\uparrow}^{\dagger},\dots,\psi_{y,\mathcal{N},\uparrow}^{\dagger},\psi_{y,1,\downarrow}^{\dagger},\dots,\psi_{y,\mathcal{N},\downarrow}^{\dagger}\right).
\tag{S10}\end{align*} $\mathcal{N}$ is the number of lattice
sites. Note that Tr$_\psi$ in Eq. \ref{Z1} represents a trace over
fermionic many-body states in Fock space. This fermionic trace can
be carried out, giving
\begin{equation*}
{\rm
Tr}\left[\prod_{n=1}^{N}\hat{B}_{n}\right]=\det\left[1+\prod_{n=1}^{N}B_{n}\right],\tag{S11}
\end{equation*}
where
$
B_{n}=e^{-\frac{1}{2}\Delta\tau K}e^{-\Delta\tau V_{n}}e^{-\frac{1}{2}\Delta\tau K}$.
 For a proof of this formula, see, e.g., Ref. (\textit{19}). 
We then arrive at the following form of the partition function:
\begin{equation*}
Z=\int
d\varphi\exp\left(-S_{\varphi}\right)\det\left[1+\prod_{n=1}^{N}B_{n}\right]+O\left(\Delta\tau^{2}\right),\label{Z2}\tag{S12}
\end{equation*}
which can be evaluated using Monte Carlo techniques, by sampling over
space-time configurations of $\vec{\varphi}_{i}\left(\tau\right)$.

\section*{Positivity of the action}

Monte Carlo sampling can be done
efficiently if the action in Eq. \ref{Z2} is non-negative. To show
that this is the case, we note that the matrix
\begin{equation}
M\left[\vec{\varphi}\right]\equiv1+\prod_{n=1}^{N}B_{n}\label{M}\tag{S13}
\end{equation}
 commutes with the following anti-unitary operator:
\begin{equation}
\mathcal{U}=is_{2}\sigma_{3}K,\tag{S14}
\end{equation}
where $\vec{s}$ are Pauli matrices which act on the spin index,
$\vec{\sigma}$ are Pauli matrices which act on the orbital ($x,y$)
index, and $K$ is the complex conjugation operator. Note that
$\mathcal{U}^{2}=-1$. Using this, one can
prove (\textit{26,27}) 
(in a similar way to the proof of
Kramers' theorem) that if $\lambda_{\alpha}$ is an eigenvalue of
$M$, $\lambda_{\alpha}^{\ast}$ is an eigenvalue also, and that if
$\lambda_{\alpha}$ is real then it is doubly degenerate. The
determinant can be written as $
\det\left[M\right]=\prod_{\alpha}\left\vert
\lambda_{\alpha}\right\vert ^{2}\geq0$. The integrand in the
partition function (Eq. \ref{Z2}) is therefore non-negative, and
can be simulated using Monte Carlo without a sign problem. Note
that there are no particular restrictions on $t_{ij}$ ({\em
e.g.\/} it does not have to be bipartite) or $\mu$. So
particle-hole symmetry or specific densities are not required.

\end{document}